\documentstyle[12pt,epsfig]{article}
\newlength{\dinwidth}
\newlength{\dinmargin}
\def\docnum#1{\hbox to \hsize{\hskip123mm\hbox{#1}\hss}}
\def\date#1{\edef\@temp{#1}\ifx\@temp\@empty\def\@temp{\today}\fi
\hbox to \hsize{\hskip123mm\hbox{\@temp}\hss}}
\def\title#1{\vskip 0.8in plus 2in\begin{center}%
{\Large\bf#1\par}\vskip1.5em\end{center}\vskip 1in}
\def\@makefnmark{\hbox{$^{\@thefnmark)}$}}
\def\author#1{
\setcounter{footnote}{0}\def\@currentlabel{}%
\begingroup\def\thefootnote{\arabic{footnote}}
\def\@makefnmark{\hbox{$^{\@thefnmark)}$}}
\global\@topnum\z@ \large\begin{center}{\lineskip.5em
\begin{tabular}[t]{c}#1\end{tabular}\par}
\end{center}\par\vskip1.5em\@thanks\endgroup}

\def\abstract{\vskip0.8in plus 3in\begin{center}{\large\bf
Abstract}\end{center}\quotation}

\begin{document}

\begin{titlepage}
\title{Maximum Relative Strangeness Content\\
 in Heavy Ion Collisions
  Around 30 A$\cdot$GeV}
\vspace{-2.6cm}
\begin{center}
\large{P. Braun-Munzinger$^a$, J. Cleymans$^b$, H. Oeschler$^c$,
and K. Redlich$^{a,d}$ }\\
\vskip 1.0 cm

{\small\it $^a$ Gesellschaft f\"ur Schwerionenforschung, D-64291
Darmstadt, Germany}
\\
{ $^b$ \small\it Department of Physics, University of Cape Town,
Rondebosch 7701, Cape Town, South Africa}\\
 {\small\it $^c$ Institut f\"ur
Kernphysik,
Darmstadt University of Technology, D-64289~Darmstadt, Germany}\\

{\small\it $^d$ Institute of Theoretical Physics, University of
Wroclaw,
PL-50204 Wroclaw, Poland.}\\
\end{center}
\vspace{-1.5cm}

\begin{abstract}
 It is shown that the ratio of strange to
non-strange particle production in relativistic heavy ion
collisions is  expected to reach a maximum   at
beam energies around 30 A$\cdot$GeV in the lab frame.
This maximum is unique to heavy ion collisions,
and has no equivalent
in elementary particle collisions.
 The appearance of the maximum
is due to the energy dependence
of the chemical freeze-out parameters and is
clearly seen as a pronounced peak in the Wroblewski factor
as a function of the incident energy as well
as in the behavior of the $K^+/\pi^+$ ratio.
Below 30 A$\cdot$GeV strange baryons
contribute strongly because of the very large value of the chemical
potential $\mu_B$.
 As the energy increases, the decrease of the
baryon chemical potential coupled with only  moderate increases in the
associated temperature causes a
decline in the
relative  number of strange baryons above energies of about
30 A$\cdot$GeV leading to very pronounced maxima in the $\Lambda/\pi^+$
and $\Xi^-/\pi^+$ ratios.\\[2cm]
\end{abstract}
\end{titlepage}

\newpage
\section{Introduction}
The experimental data from heavy ion collisions show that
the $K^+/\pi^+$ ratio is larger at BNL-AGS energies than at the
highest CERN-SPS energies
[1-6] and even at RHIC \cite{harris}.
 This behavior is of
particular interest as it could signal the appearance of new
dynamics for strangeness production in high energy collisions. It
was  even conjectured~\cite{gazdzicki}
 that this property could
indicate   an energy    threshold  for  quark-gluon plasma
formation in relativistic heavy ion collisions. In this paper we
analyze the energy dependence of strange to non-strange
particle  ratios in the framework of a hadronic
statistical model. The statistical approach has been
  very successful in describing particle yields
from low energies starting with SIS and AGS all
 the way up to SPS [9-16] and
 RHIC \cite{redlich,xu,pbm3} energies.
In particular it was found that, in the whole  energy range, the
hadronic yields  observed in heavy ion collisions resemble those
of a population in chemical equilibrium along a unified freeze-out
curve determined by the condition of fixed energy/particle
$\simeq$ 1 GeV \cite{prl}. The chemical freeze-out curve provides
a relation between the temperature $T$ and the baryon chemical
potential  $\mu_B$ when the particle composition of the system is
frozen in.

In this paper we use   the energy dependence of the thermal
parameters at chemical freeze-out as determined from the analysis
of [10-16] to show that the ratio of strange to non-strange
particle multiplicities reaches  a maximum at approximately  30
A$\cdot$GeV lab energy.
 Compared to other $q\bar{q}$
pairs, the relative strangeness content actually declines by
almost a third between  30 A$\cdot$GeV and RHIC energies.
We emphasize that this maximum is unique to heavy
ion collisions and does not occur
 in the collisions of elementary particles like p-p, p-$\bar{\rm p}$
or e$^+$-e$^-$.
%
%
%

The values of the baryon chemical potentials
and temperatures
at different collision energies are shown in Fig. 1.
The values of $\mu_B$ at SIS, AGS and SPS energies were  taken
 from a detailed analysis~\cite{CLE99,becattini} of the
data on  particle multiplicities  extrapolated to
full phase space
 while the value of $\mu_B$ at RHIC was obtained from data taken at
 mid-rapidity~\cite{pbm3}.
The energy dependence
can be parametrized phenomenologically as
\begin{equation}
\mu_B(s) \simeq {a~{\mathrm{}}\over(1+ \sqrt{s}/b)}
\end{equation}
where  $a\simeq 1.27$ GeV and $b\simeq 4.3$ GeV. The results of
this parametrization are  shown by the full line in the upper part of Fig. 1.
At  freeze-out the chemical potential  is  related to the
temperature  via the phenomenological   condition of fixed
energy per hadron~\cite{prl}, namely
\begin{equation}
\left< E\right>/\left< N\right> \simeq 1~\mathrm{GeV} .
\end{equation}
Applying Eq.~(1) together with Eq.~(2)  leads to  the energy
dependence of the temperature shown  by the full line in lower part of Fig.~1
together with the values obtained in~\cite{becattini,pbm3}.

Both the temperature and the chemical potential exhibit a strong
variation with energy. The GSI/SIS results have the lowest
freeze-out temperatures and the highest baryon chemical
potentials. As the beam energy is increased a clear shift towards
higher $T$ and lower $\mu_B$ occurs. Above AGS energies, the
 temperature exhibits only a
moderate change and converges to
its maximal value in the range of 160 to 180 MeV. This value is
remarkably close to the critical temperature
 $T_c\sim 170\pm 8$ MeV extracted from the lattice calculations of QCD
\cite{karsch}.

In the statistical approach,  the energy dependence of
the basic thermal parameters determines  the energy
dependence of many relevant  observables, for instance
particle yields. For strange
particle  production
this determines the ratios of
 strange to non-strange particle
multiplicities as well as the Wroblewski
factor~\cite{wroblewski}
defined as
\begin{equation}
\lambda_s \equiv {2\bigl<s\bar{s}\bigr>\over
\bigl<u\bar{u}\bigr> + \bigl<d\bar{d}\bigr>}
\end{equation}
where the  quantities in angular brackets refer to the number of
newly formed quark-antiquark pairs, i.e., it excludes all
quarks that were present in the target and projectile.

Applying the statistical model to particle production in heavy ion
collisions calls for the  use of the canonical ensemble
to treat the number of strange particles
particularly  for data in the energy range
from SIS up to AGS \cite{CLE99,ko}. For these energies, the
 number of strange particles per event is so small that the description of
strangeness conservation on the average, as contained in the grand
canonical approach is not adequate.
 The  exact conservation of quantum numbers  in
relativistic statistical mechanics has been well established for
some time now \cite{hagedorn, esko}. In the following we present
 a new analytical expression of the
canonical partition function and the corresponding multi-strange
particle multiplicities. This is quantitatively equivalent to
previous results~\cite{esko,hamieh}, however, the result  can now
be presented in  a compact format well suited for numerical
evaluation.

\section{Canonical strange particle multiplicities}

The canonical partition function of a  hadronic resonance gas
constrained by the strangeness neutrality condition reads
\cite{esko},

\begin{equation}
Z^C_{S=0}={1\over {2\pi}}
       \int_{-\pi}^{\pi}
    d\phi~ \exp{\left(\sum_{n=- 3}^3S_ne^{in\phi}\right)},
\end{equation}
where $S_n= V\sum_k Z_k^1$ with $V$ being the volume  and the sum
is  over all particles and resonances carrying strangeness $n$.
For a particle of  mass $m_k$, with spin-isospin degeneracy factor
$g_k$, carrying  baryon number $B_k$ and charge $Q_k$, with the
corresponding chemical potentials $\mu_B$ and $\mu_Q$,
 the one-particle partition function
is given, in Boltzmann approximation, by
\begin{equation}
 Z_k^1\equiv {{g_k}\over {2\pi^2}}
m_k^2TK_2\left({{m_k}\over T}\right)\exp (B_k\mu_B+Q_k\mu_Q) . \label{e:zk1}
\end{equation}

The integral representation of the partition function in Eq.~(4)
is not convenient for a numerical analysis as the integrand  is a
strongly oscillating function. However, the $\phi$ integration in
Eq.~(4)  can be done exactly. Indeed, rewriting  Eq.~(4) as

\begin{equation}
Z^C_{S=0}={1\over {2\pi}}e^{S_0}
       \int_{-\pi}^{\pi}
                   d\phi~ \prod_{n=1}^3
\exp{\left[{{x_s}\over 2}
\left(a_ne^{in\phi}+ a_n^{-1}e^{-in\phi}\right)\right]}, \label{eq1}
\end{equation}
and using the relation    $e^{{\rho\over 2}(t+{1\over
t})}=\sum_{-\infty}^{+\infty}t^mI_m(\rho)$ one obtains, after
integrating,  the partition function in a form  free of oscillating
terms
\begin{equation}
Z^C_{S=0}=e^{S_0}
\sum_{n=-\infty}^{\infty}\sum_{p=-\infty}^{\infty} a_{3}^{p}
a_{2}^{n} a_{1}^{{-2n-3p}} I_n(x_2) I_p(x_3) I_{-2n-3p}(x_1)
 , \label{eq2}
\end{equation}
where
\begin{eqnarray}
a_i&=& \sqrt{{S_i}/{S_{-i}}}\\
x_i &=& 2\sqrt{S_iS_{-i}}
\end{eqnarray}
and $I_i$ are modified Bessel functions. The contribution of
thermal pions to $S_0$ is calculated using Bose-Einstein
statistics.

The expression for the  particle density, $n_i$, can be obtained
from the partition function Eq.~(3) by following the standard
method \cite{esko}. For a particle $i$ having strangeness $s$ the
result is
\begin{equation}
n_{i}={{Z^1_{i}}\over {Z_{S=0}^C}}
\sum_{n=-\infty}^{\infty}\sum_{p=-\infty}^{\infty} a_{3}^{p}
a_{2}^{n}
 a_{1}^{{-2n-3p- s}} I_n(x_2) I_p(x_3) I_{-2n-3p- s}(x_1)
 . \label{eq5}
\end{equation}
The leading term in $x_3$ corresponds to the
approximate form used in~\cite{hamieh}
to study the centrality
dependence of strangeness production  in Pb-Pb collisions at SPS energies.

 It can be   verified  that, in the limit of large $x_i$,
the above formulae  coincide with the grand canonical result. In the
opposite  limit, that of small $x_i$, the equilibrium density of
(multi)strange particles is strongly suppressed relative to its
grand canonical value. The above equation can therefore be applied
both at SIS energies where the number of strange particles
produced per event is as small as $10^{-2}$, and at SPS
where  over a hundred strange particles are  produced on average in each
event.
\section{Results}

The theoretical results of the previous section combined with the energy
dependence of the thermal parameters shown in  Fig.~1 provide a
basis for the study of the energy dependence of strangeness production
in heavy ion collisions. Of particular interest are  the ratios of
strange to non-strange particle multiplicities as well as the
relative strangeness content of the system as expressed by the
Wroblewski factor \cite{wroblewski}.
 The thermal model calculations for Au-Au and
Pb-Pb collisions  are performed using
a canonical correlation volume given by a radius of $\sim 7$ fm.
 This radius
could vary with energy but for simplicity this was not taken into account
in our calculations. At a later stage, when more
data become available,
this approximation will of course have to be reconsidered.
Furthermore, we assume strangeness to be in complete equilibrium,
that is, the  strangeness saturation factor is taken as  $\gamma_s=1$.
When more data will become available, this point might be refined.

We turn our attention first
to  the energy dependence of the Wroblewski ratio
The quark content used in this ratio is determined at the moment
 of {\it {chemical freeze-out}}, i.e.
from the hadrons and especially, hadronic
resonances, before they decay. 
This ratio is thus not an easily measurable  observable
unless one can reconstruct all resonances from the final-state
particles.
 The results are shown in Fig.~2 as a function of
invariant energy $\sqrt{s}$. The values calculated from the
experimental data at chemical freeze-out  in central A-A
collisions have been taken from
reference~\cite{becattini}\footnote{ There the statistical model
was fitted with an extra parameter $\gamma_s$ to account for
possible chemical undersaturation of strangeness. At the SPS,
$\gamma_s\simeq 0.7$ give the best agreement with 4$\pi$ data.}.
The values of $\lambda_s$ were extracted from fully integrated
4$\pi$ data with the exception of the result form RHIC where
particle ratios measured at mid-rapidity \cite{harris} were used.
The solid line in Fig.~2 describes the statistical model
calculations in complete equilibrium along the unified freeze-out
curve~\cite{prl} and with the energy dependent thermal parameters
presented here. From Fig.~2 we conclude that around 30 $A\cdot$GeV
lab energy the relative strangeness content in heavy ion
collisions reaches a
 clear and well pronounced maximum.
The Wroblewski factor  decreases towards higher incident energies
and reaches a limiting value of about 0.43.

 The
appearance of the maximum can be traced  to the specific
dependence of $\mu_B$ on
the beam energy. 
In Fig.~2  we also show the results  for  $\lambda_s$ calculated
under the assumption that only the temperature  varies with
collision energy but the baryon chemical potential is kept fixed
at zero. In this case the Wroblewski factor  is indeed seen to be
a monotonic function of energy. The assumption of vanishing net
baryon density is close to the prevailing situation in e.g.
p-$\bar{\rm{p}}$ and e$^+$-e$^-$ collisions. In Fig.~2 the results
for $\lambda_s$ extracted from the data in p-p, p-$\bar {\rm p}$ and
e$^+$-e$^-$ are also included~\cite{becattini}. The dashed line
represents results with $\mu_B= 0$ and a  radius of 1.2 fm. There
are two important differences in the behavior of $\lambda_s$ in
elementary compared to heavy ion collisions. Firstly, the
strangeness content is smaller by a factor of two.
In elementary collisions particle multiplicities follow
the values given by the canonical ensemble with radius 1.2 fm
 whereas in A-A collisions there is a transition from
canonical to grand canonical behavior. Secondly, there is no
significant maximum in the behavior of $\lambda_s$ in elementary
collisions
 due to the  vanishingly small baryon density in the p-$\bar{\rm{p}}$ and
e$^+$-e$^-$ systems.

 The position of the
maximum is further clarified in Fig.~\ref{Wrob_T_MUB} which shows
values of constant $\lambda_s$  in the $T-\mu_B$
plane. As expected $\lambda_s$ rises with increasing $T$ for fixed
$\mu_B$.
Following the chemical freeze-out curve, shown as a thick
full line in
Fig.~\ref{Wrob_T_MUB}, one can see that
 $\lambda_s$ rises quickly from SIS to AGS energies,
then reaches  a maximum around $\mu_B\approx 500$ MeV
and $T\approx 130$ MeV.
These freeze-out parameters correspond to
30 GeV lab energy. At higher incident
energies the increase in $T$ becomes negligible but $\mu_B$ keeps
on decreasing and as a consequence $\lambda_s$ also decreases.

 The importance of finite baryon density on the
behavior of $\lambda_s$ is demonstrated in  Fig.~4  showing
 separately the  contributions to $\left<s\bar{s}\right>$
coming from strange
baryons, from strange mesons and from hidden strangeness, i.e.,
from hadrons  like $\phi$ and $\eta$.
 As can be seen in Fig.~4,
the origin of the maximum in the Wroblewski
 ratio can be traced  to the contribution
of strange baryons.
Even strange mesons exhibit a broad maximum. This is due to the
presence of associated production of e.g.~kaons together with
hyperons. This channel dominates at low $\sqrt{s}$ and loses
importance at  high incident energies.
%

Next we study how the behavior of the Wroblewski ratio is reflected in
specific particle yields.
Of particular interest is the $K$ yield which at high
energies is responsible for  almost 80$\%$ of the
total strangeness production while
at lower energies this contribution decreases to 50$\%$,
due to the associated production with hyperons.

 The energy dependence of the
$K^+/\pi^+$ ratio measured at midrapidity
is shown in Fig. 5.
The model  gives an excellent description of the data, showing
a broad maximum at the same energy as the one
seen in the Wroblewski factor.
In general, of course, statistical-model calculations
 should be compared with
4$\pi$-integrated results since strangeness does not have to be
conserved in a limited portion of phase space. A drop in this
ratio for 4$\pi$ yields has been reported from preliminary results
of the NA49 collaboration  at 158 AGeV~\cite{blume}. This decrease
is, however, not reproduced by the statistical model
 without further modifications, e.g.~by introducing an additional
parameter $\gamma_s\sim 0.7$ \cite{becattini}.
This point might be clearer when final data, also at other beam
energies  will become available.

The appearance of the maximum in the relative strangeness contribution
becomes also obvious when considering ratios which are more
sensitive to the baryon chemical potential. Figure~6 shows the energy
dependence of the $\Lambda /\pi^+$,  $\Xi^- /\pi^+$ and
the $\Omega/\pi^+$ ratios.
As can be seen from the figure  there is a very clear pronounced
maximum especially in the $\Lambda/\pi^+$ ratio. The relative enhancement of
$\Lambda$ is stronger then that of $\Xi^-$ or $\Omega^-$.
There is also a shift of the maximum  to higher energies for particles with
increasing strangeness quantum number.
These  differences appear  as a consequence  of the
enhanced strangeness content of the particles which suppresses
the dependence of the corresponding ratio on $\mu_B$.
\section{Conclusions}
In conclusion, using the energy dependence of thermal parameters
determined from an analysis of available data,  we have  shown
that the statistical model description of relativistic heavy ion
collisions predicts that the yields of strange to nonstrange
particles reaches  a well defined maximum near 30 GeV lab energy.
It is demonstrated that this maximum is due to the specific shape
of the freeze-out curve in the $T-\mu_B$ plane. In particular,  a
very steep decrease of the baryon chemical potential with
increasing energy causes a corresponding decline of relative
strangeness content in thermal systems created in heavy ion
collisions above lab energies of 30 GeV. The saturation in $T$,
necessary for this result, might be connected to the fact that
hadronic temperatures cannot exceed the critical temperature
$T_c\simeq$ 170 MeV for the phase transition to the QGP as found
in solutions of QCD on the lattice.

The maximum in the relative strangeness content is unique to heavy
ion collisions:  there is no equivalent behavior
 in the collisions of elementary particles.
It has already been observed in the energy dependence of the
$K^+/\pi^+$ ratio, and  is predicted to be even more pronounced in
the $\Lambda/\pi^+$ ratio. The maxima in the $\Xi^-/\pi^+$ and
$\Omega^-/\pi^+$ ratios are expected to occur at a slightly higher
beam energy and to be less pronounced.

It is thus clear that in order to study collective effects related
to strangeness production, one interesting area is the energy
region around 30 A$\cdot$GeV. Such a program is being planned for
the near future at GSI.
\section*{Acknowledgments}
On of us (K.R.)  acknowledges a stimulating discussions with M.
Ga\`zdzicki and  B. Friman as well as the partial support of the
Polish Committee for Scientific Research (KBN-2P03B 03018). J.C.
acknowledges partial financial support by the URC of the
University of Cape Town. We also acknowledge stimulating
discussions with the members of the GSI theory division.
%
%

%
%
%
%
%
\newpage
\begin{figure}
\hspace*{1cm}
\includegraphics[width=.9\textwidth]{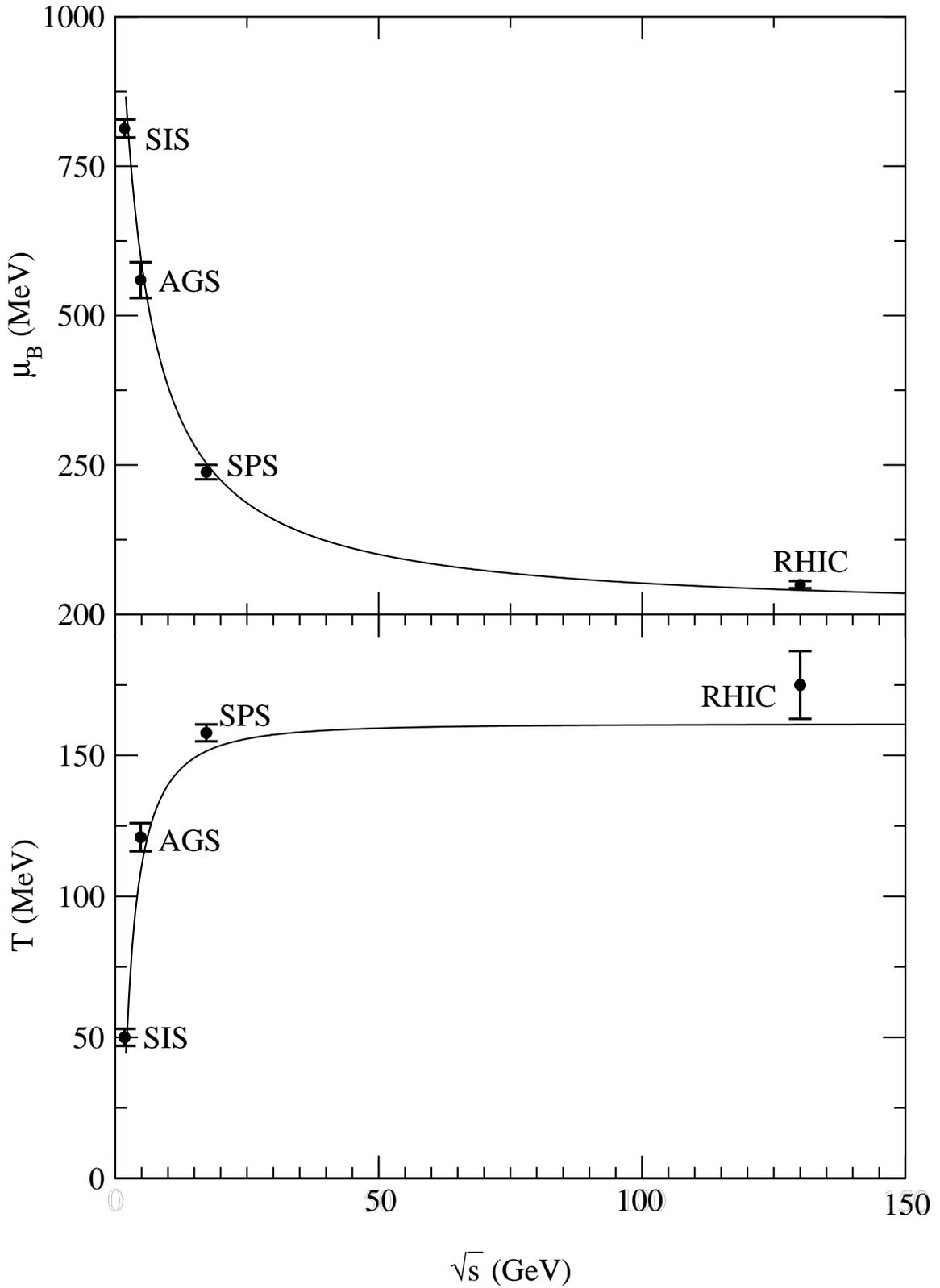}
\caption{{ Behavior of the freeze-out baryon chemical potential
$\mu_B$ (upper curve) and the temperature $T$ (lower curve) as a
function of energy. The temperature  $T$ as a function of beam
energy is determined from the  unified freezeout condition
$\left<E\right>/\left<N\right>$ = 1 GeV \cite{prl}. }}
\label{mubT_e}
\end{figure}
\newpage
\begin{figure}
\includegraphics[width=0.9\textwidth]{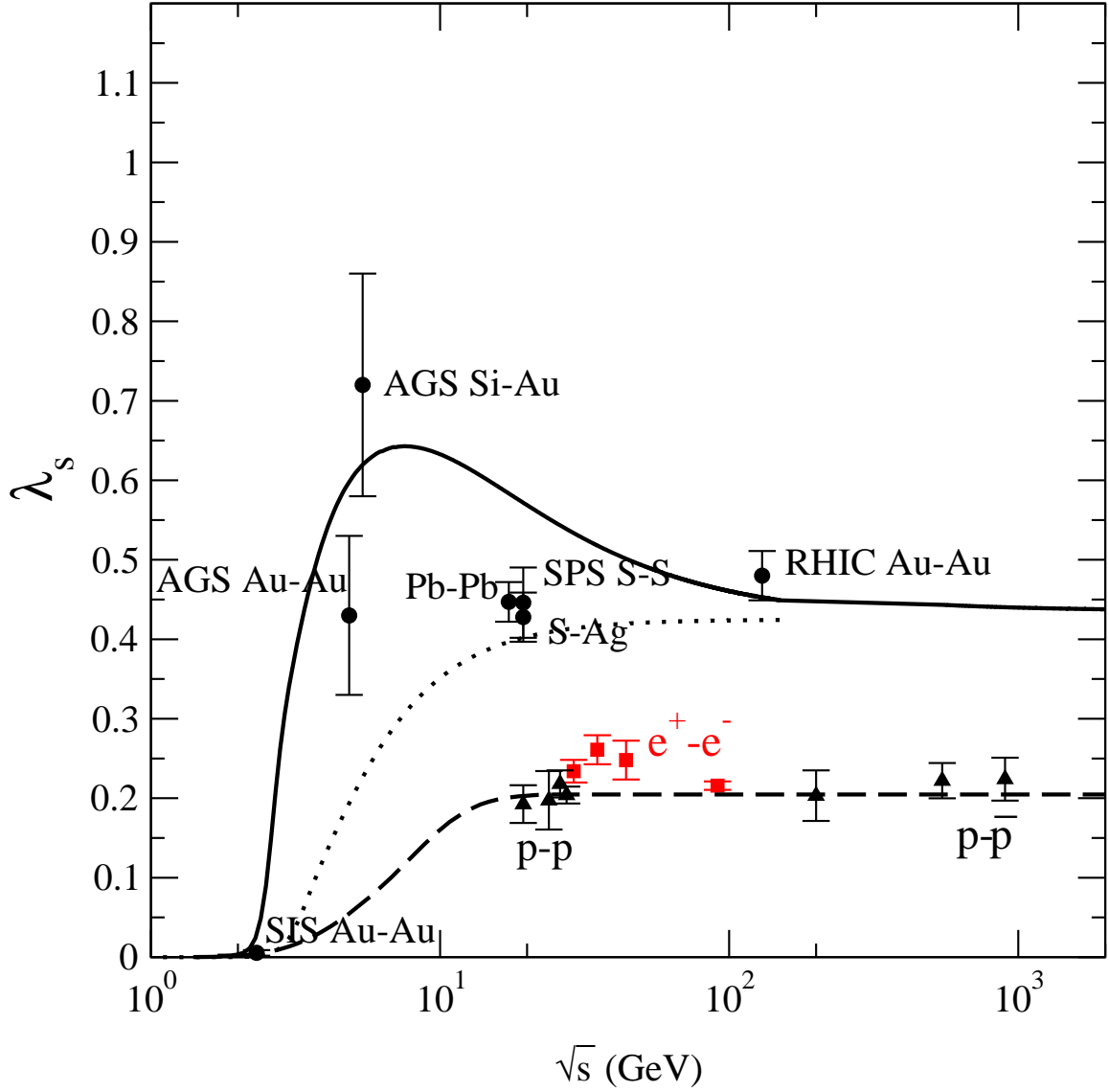}
\caption{{ The Wroblewski ratio $\lambda_s$ (for definition see
text) as a function of $\sqrt s$. The thick  solid line has been
calculated using the freeze-out values of the temperature and the
baryon chemical potential. The dotted line has been calculated
using $\mu_B=0$ and only varying $T$. The dashed line has been
calculated using a radius of 1.2 fm, keeping $\mu_B$=0 and taking
the energy dependence of the temperature as determined previously.
All calculations are performed using strangeness saturation
$\gamma_s=1$.}}
\label{Wrob_sqrts}
\end{figure}
\newpage
\begin{figure}
\includegraphics[width=0.9\textwidth]{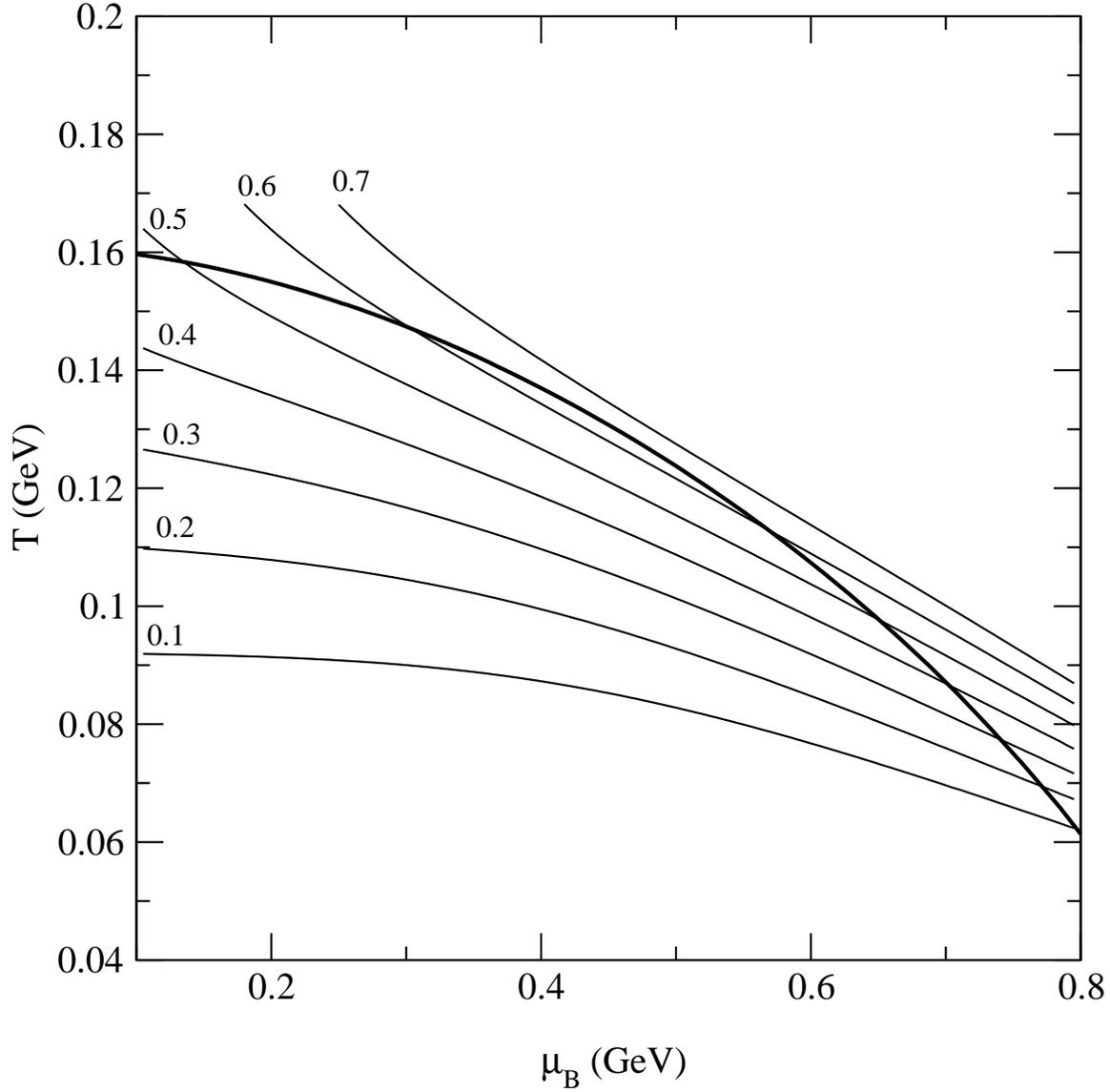}
\caption{{ Lines of constant Wroblewski factor $\lambda_s$ (for
definition  see text) in the $T-\mu_B$ plane (thin solid lines)
together with the freeze-out curve (thick solid line)~\cite{prl}.
}} \label{Wrob_T_MUB}
\end{figure}
\newpage
\begin{figure}
\hspace*{1cm}
\includegraphics[width=.9\textwidth]{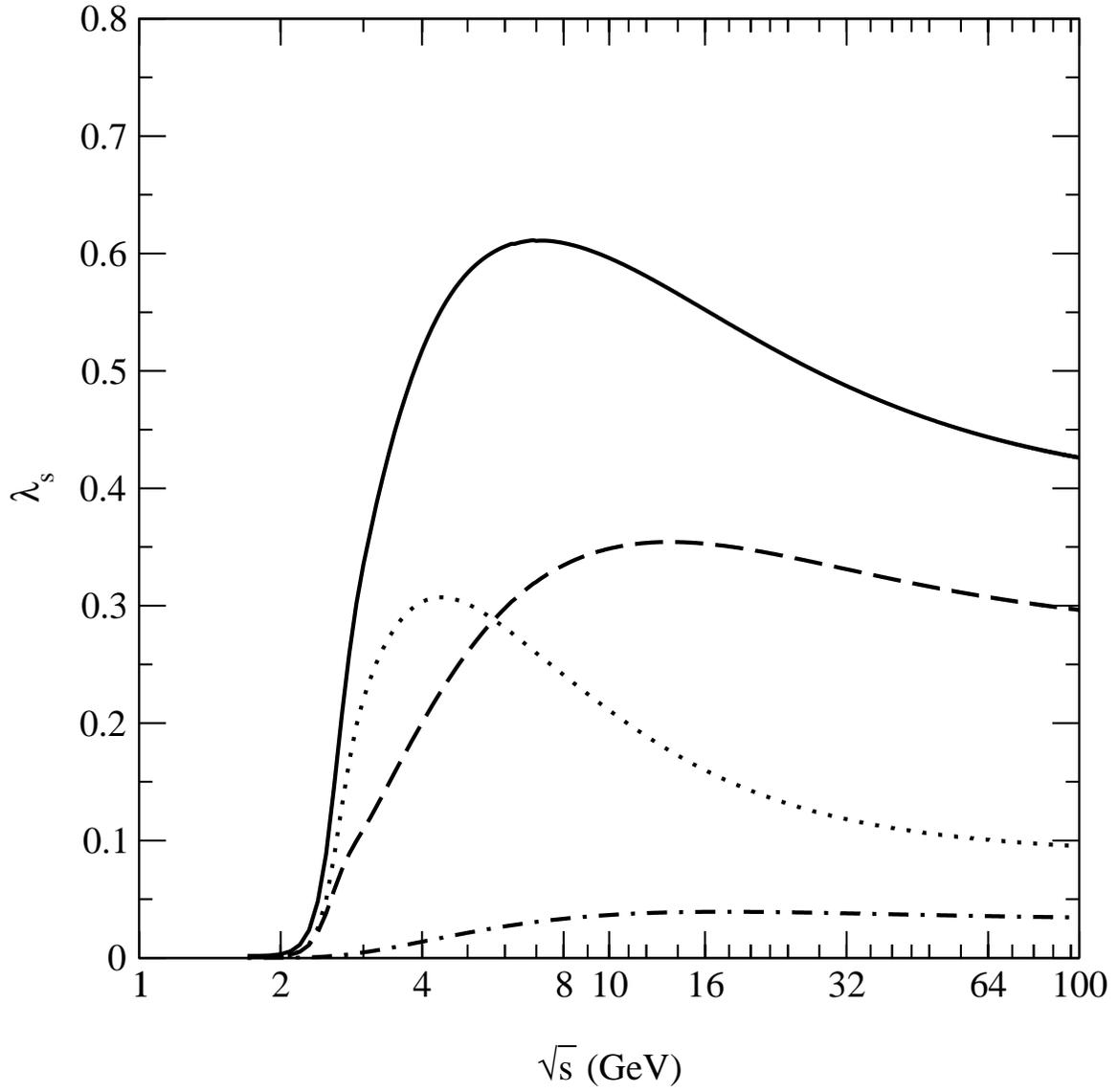}
\caption{ Contributions to the Wroblewski factor (for definition
see text) from strange baryons (dotted line), strange mesons
(dashed line) and mesons with
 hidden strangeness (dash-dotted line).
The sum of all contributions is given by the full line}
\label{composition}
\end{figure}
\newpage
\begin{figure}
\hspace*{1cm}
\includegraphics[width=.9\textwidth]{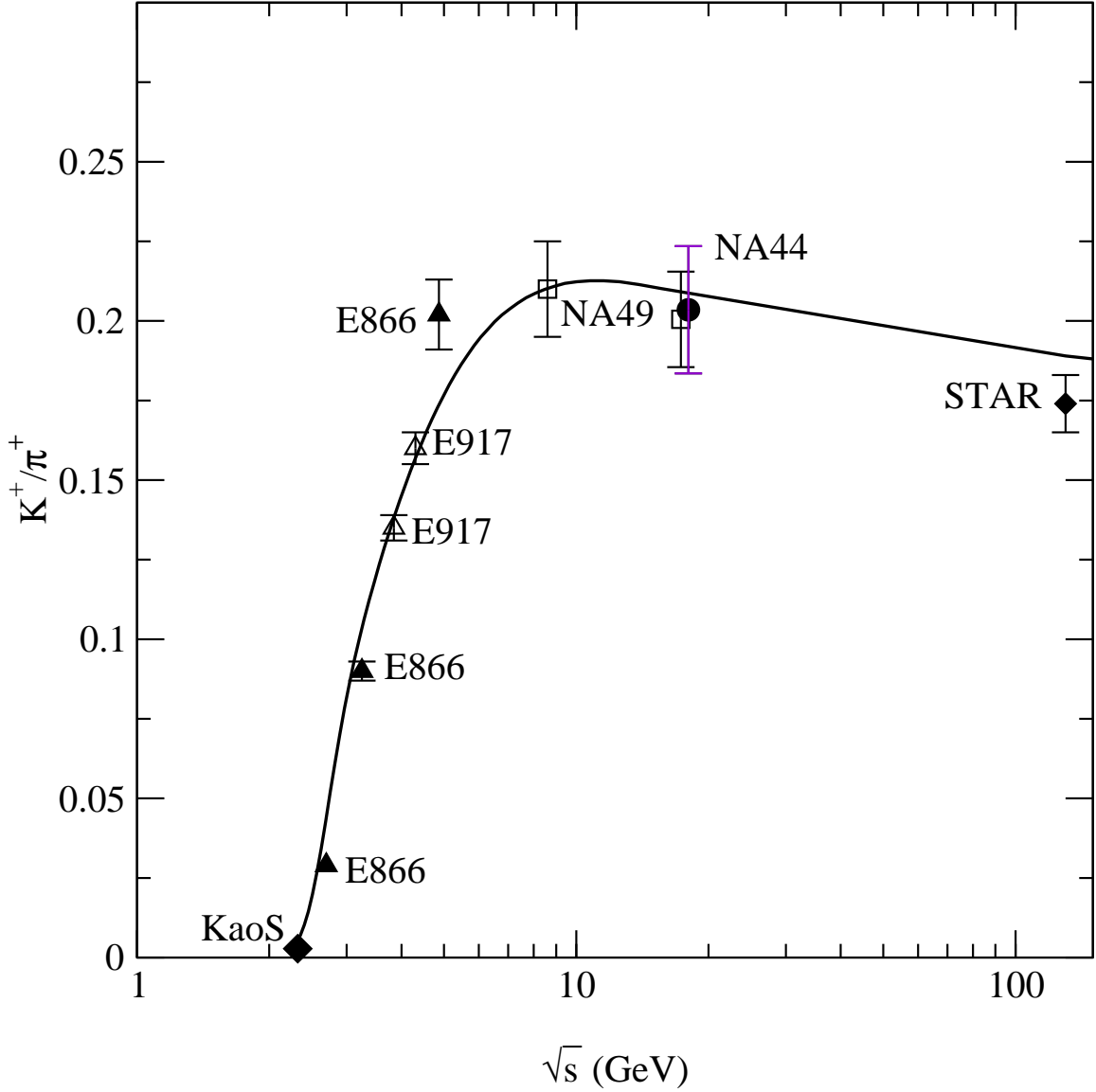}
\caption{{ $K^+/\pi^+$ ratio obtained around midrapidity as a
function of $\sqrt s$ from the various experiments. For the
references for all data points see \cite{blume,redlich} . The full
line shows the results of the statistical model in complete
equilibrium. The value at RHIC  was estimated using results from
the STAR collaboration on the $K^-/\pi^-$ and $K^+/K^-$ ratios,
assuming $\pi^-/\pi^+=1.007$.}} \label{KP_PI_ratio}
\end{figure}
\begin{figure}
\hspace*{1cm}
\includegraphics[width=.9\textwidth]{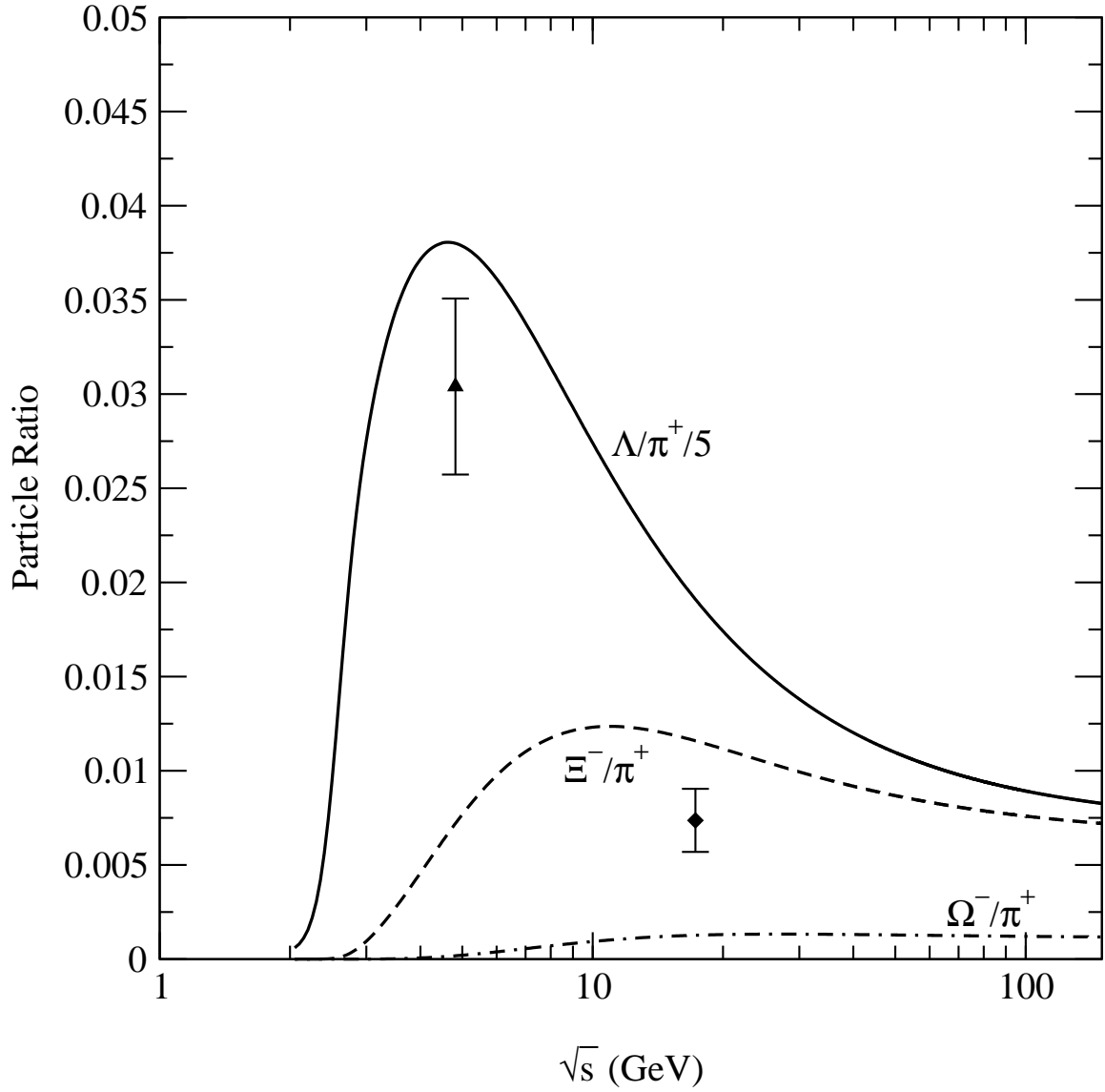}
\caption{{ Prediction for the $\Lambda/\pi^+$ (note the factor 5),
$\Xi^-/\pi^+$ and the $\Omega^-/\pi^+$  ratios as a function of
$\sqrt s$. For  compilation of data see \cite{becattini}.}}
\label{figure6}
\end{figure}

\begin{thebibliography}{10}
\bibitem{E802} L.~Ahle et al., (E802 Collaboration),
Phys. Rev.{ C60} (1999) 044904; L.~Ahle et al., E866/E917
Collaboration, Phys. Lett.~{ B476} (2000) 1.

\bibitem{dunlop} J. C. Dunlop and C. A. Ogilvie, Phys. Rev. C61
(2000) 031901 and references therein;\\
 C. A. Ogilvie. Talk
presented at QM2001, Stony Brook, January 2001, Nucl. Phys. A
(2001) (in print).

%
\bibitem{blume} Ch. Blume, NA49 Collaboration. Talk
presented at QM2001, Stony Brook, January 2001, Nucl. Phys. A
(2001) (in print).

\bibitem{bearden} I. Bearden, NA44 Collaboration, Phys. Lett. B471
(1999) 6.

%
\bibitem{cleymans1} J. Cleymans and K. Redlich,  Phys. Rev.~{C60} (1999) 054908.
%

\bibitem{redlich} K.~Redlich,  hep-ph/0105104. Talk presented at
QM2001, Stony Brook, January 2001,  Nucl. Phys. A (2001) (in
print).

\bibitem{harris} J. Harris,
STAR Collaboration. Talk presented at QM2001, Stony Brook, January
2001, Nucl. Phys. A (2001) (in print).

\bibitem{gazdzicki} M. Ga\`zdzicki and M. Gorenstein,
Acta Phys. Pol.  B30 (1999) 2705;  M. Ga\,zdzicki and D. Rohrich,
Z. Phys. C71 (1996) 55.
\bibitem{satz} J.~Cleymans and H.~Satz, Z.~Phys.~{ C57} (1993) 135.

\bibitem{pbm1}P. Braun-Munzinger, J. Stachel, J.P. Wessels
and N. Xu, Phys. Lett.~{ B344} (1995) 43; B365 (1996) 1.
\bibitem{pbm2} P. Braun-Munzinger, I.~Heppe, J.~Stachel, Phys. Lett. { B465}
 (1999) 15.
%

\bibitem{stachel} P. Braun-Munzinger and J. Stachel, Nucl. Phys. A606 (1996) 320;
J. Stachel, Nucl. Phys.  A654 (1999)  119c.

\bibitem{prl} J. Cleymans and K. Redlich, Phys. Rev.
Lett. { 81} (1998) 5284.

\bibitem{CLE99} J. Cleymans, H. Oeschler and K. Redlich, Phys. Rev.~{ C59}
(1999) 1663; Phys. Lett.~{ B485} (2000) 27.

\bibitem{becattini} F. Becattini, J. Cleymans, A. Ker\"anen, E. Suhonen and K. Redlich, hep-ph/0002267, Phys.
Rev. C (in print).

\bibitem{oeschler}
 H.~Oeschler, J.~Phys.~G: Nucl.~Part.~Phys. { 27} (2001) 1.
%
\bibitem{xu}
N. Xu,  nucl-ex/0104021. Talk presented at QM2001, Stony Brook,
January 2001, Nucl. Phys. A (2001) (in print).

\bibitem{pbm3} P. Braun-Munzinger, D.J. Magestro, K. Redlich
 and J. Stachel, hep-ph/0105223.

\bibitem{karsch} F. Karsch, Nucl.Phys. B (Proc. Suppl.) 83-84
(2000) 14.

\bibitem{wroblewski} A. Wroblewski, Acta Physica Polonica~{ B16} (1985) 379.



\bibitem{ko}  C.M. Ko, V. Koch, Z. Lin, K. Redlich,
M. Stephanov and X.N. Wang,  Phys. Rev. Lett. 86 (2001) 5438.

\bibitem{hagedorn}R. Hagedorn, CERN Rep. 71 (1971); E.
Shuryak, Phys. Lett. B42 (1972) 357; J. Rafelski,  Phys. Lett. B97
(1980) 279; K. Redlich and L. Turko, Z. Phys. { C5 } (1980) 201;
R. Hagedorn and K. Redlich, Z. Phys. { C27} (1985) 541; F.
Becattini, Z. Phys. C69 (1996) 485; F. Becattini and U. Heinz, Z.
Phys. C76 (1997) 269.

\bibitem{esko} J. Cleymans, K. Redlich and E. Suhonen, Z. Phys. C76 (1997)  269.

\bibitem{hamieh}J. S. Hamieh, K. Redlich and A. Tounsi,
Phys. Lett. { B486} (2000) 61.
\end{thebibliography}
\end{document}